\documentclass{article}

\usepackage{PRIMEarxiv}

\usepackage[utf8]{inputenc} % allow utf-8 input
\usepackage[T1]{fontenc}    % use 8-bit T1 fonts
\usepackage{hyperref}       % hyperlinks
\usepackage{url}            % simple URL typesetting
\usepackage{booktabs}       % professional-quality tables
\usepackage{amsfonts}       % blackboard math symbols
\usepackage{nicefrac}       % compact symbols for 1/2, etc.
\usepackage{microtype}      % microtypography
\usepackage{lipsum}
\usepackage{fancyhdr}       % header
\usepackage{graphicx}       % graphics

\usepackage{subcaption} % 推荐用于子图
\usepackage{caption}

\usepackage[ruled,vlined]{algorithm2e}
\usepackage{algorithmic}
\usepackage{amsmath, amssymb}
\graphicspath{{media/}}     % organize your images and other figures under media/ folder

\SetKw{KwBy}{by}

\title{W4A16 Mixed-Precision Matrix Multiplication on Decoupled Architecture: Kernel Design and Memory Bottleneck Analysis for Ascend NPUs
}

\author{
  Yuanhong He \\
    PKU-Changsha Institute of \\
Computing and Digital Economy \\
  Changsha, China\\
  \texttt{yuanhong\_he@163.com} \\
   \And
  Peiyu Niu \\
  PKU-Changsha Institute of\\
Computing and Digital Economy\\
  Changsha, China\\
  \texttt{niupeiyu@icode.pku.edu.cn} \\  
   \And
  Jun Chen \\
  PKU-Changsha Institute of\\
Computing and Digital Economy\\
  Changsha, China\\
  \texttt{chenjun@icode.pku.edu.cn} \\  
  \And
Chenchen Zhang \\
Peking University\\
Beijing, China\\
  \texttt{zhangchenchen@stu.pku.edu.cn} \\  
  \And
  Chao Yang \\
  PKU-Changsha Institute of\\
Computing and Digital Economy\\
  Changsha, China\\
  Peking University\\
  Beijing, China\\
  \texttt{chao\_yang@pku.edu.cn} \\
}

\begin{document}
\maketitle

\begin{abstract}
As Large Language Models (LLMs) scale, weight-only quantization (W4A16: 4-bit weights, 16-bit activations) becomes critical for reducing memory footprint with minimal accuracy loss. However, its efficient deployment on Huawei's Ascend 910 Neural Processing Unit (NPU) is challenging due to limited native mixed-precision support and the accelerator's decoupled compute architecture.  To enable quantization on such architecture, we present the first practical W4A16 matrix multiplication kernel tailored for the Ascend 910 NPU. Our design leverages vector cores for on-the-fly INT4-to-FP16 dequantization, cube cores for high-throughput GEMM, and Split-K parallelization to mitigate memory latency. Performance evaluations across diverse matrix shapes and batch sizes show our method outperforms data-parallel approaches when K $\gg$ N, a typical scenario in LLM decoding. Specially, our method can achieve a speedup ranging from $1.01\times$ to $1.74\times$. In addition, our profile reveals the primary bottleneck is not dequantization computation itself, but extra global memory transfer for the weight, making W4A16 only reaching a maximum speedup of $1.48\times$ over native FP16$\times$FP16 matrix multiplication in PyTorch. In the long run, our method lays a solid foundation and provides insightful views for the efficient deployment of quantized large language models on various domain-specific accelerators.
\end{abstract}

% keywords can be removed

\section{Introduction}

Quantization~\cite{jacob2018quantizationgoogle} has emerged as a pivotal technique for deploying large language models (LLMs)~\cite{touvron2023llama, yang2025qwen3, liu2024deepseek, chen2025pangu, zeng2025glm} on resource-constrained hardware. Recent studies, such as SINQ~\cite{muller2025sinq} and QQQ~\cite{zhang2024qqq}, have achieved significant reductions in memory footprint, bandwidth requirements, and computational cost without incurring substantial losses in model accuracy. Among various quantization schemes, weight-only strategies—wherein weights are quantized to fixed-point integers while activations remain in floating-point format—have garnered growing attention due to their capability to balance efficiency and fidelity. In particular, the W4A16 configuration (4-bit weights and 16-bit activations)~\cite{frantar2022gptq, lin2024awq} has demonstrated remarkable potential: it retains high activation precision to preserve the dynamic range, while aggressively compressing weights to minimize storage and memory access overhead.
However, unlocking the full potential of W4A16 quantization necessitates specialized kernel implementations that can efficiently unpack and dequantize 4-bit weights into FP16/BF16 formats prior to matrix multiplication. General Matrix Multiplication (GEMM) routines are ill-suited for such preprocessing-intensive, irregular data layouts and often fail to fuse dequantization with computation. This deficiency results in suboptimal memory bandwidth utilization and redundant data movement—challenges that are particularly pronounced on domain-specific architectures such as Huawei's Ascend NPU~\cite{CloudMatrix384}. While the cube units of the Ascend NPU are highly optimized for native GEMM operations, they lack tailored support for the additional memory-bound dequantization steps required by 4-bit weight formats.

To address this gap, we present the first practical implementation of a W4A16 matrix multiplication kernel customized for Huawei's Ascend NPU. Notably, while NVIDIA platforms benefit from highly optimized open-source libraries (e.g., Bitblas~\cite{ladder-osdi24} and Marlin~\cite{frantar2025marlin}) that support low-bit quantized inference, the Ascend software stack currently provides no native kernel support for mixed-precision matrix multiplication. Our implementation integrates on-the-fly weight dequantization with compute scheduling, enabling the direct execution of W4A16 mixed-precision matrix multiplication kernels on Ascend NPUs.

\section{Background}
\label{sec:headings}
\label{sec:background}

\subsection{Quantization in Deep Neural Networks}
Quantization is a technique that maps high-precision floating-point parameters to lower-bit integer representations, primarily to reduce model size and computational cost. Among various approaches, weight-only quantization has emerged as a highly effective strategy for compressing and accelerating Large Language Models. In this paradigm, only the model weights are quantized—often to 4 bits or lower—while activations remain in full precision (e.g., FP16). This design preserves model accuracy more effectively than full quantization by avoiding error propagation through low-precision intermediate computations. Consequently, weight-only methods are particularly suitable for deployment on memory-constrained devices, as they dramatically reduce the model's memory footprint.

A widely used formulation for weight quantization is uniform affine quantization\cite{jacob2018quantizationgoogle}, defined as:
\begin{equation}
x_q = \text{round}\left( \frac{x}{s} \right) + z
\end{equation}
where \(x\) is the original value, \(x_q\) the quantized integer, \(s\) the scale factor, and \(z\) the zero-point offset. For symmetric quantization (common in weights), \(z = 0\).

\subsection{Motivation and Challenges}
W4A16\cite{frantar2022gptq,lin2024awq} quantization (4-bit weights, 16-bit activations) offers an trade-off between model efficiency and accuracy. By storing weights in 4-bit integers, it reduces model size by approximately 4× compared to FP16 representations. However, a key implementation challenge remains: mainstream hardware—including GPUs and NPUs—lacks native support for mixed-precision arithmetic operations. In practice, 4-bit values must be packed into larger data types such as INT32 for storage and transfer, then dequant prior to computation. This dequant process introduces additional overhead and can become a notable performance bottleneck unless carefully optimized through efficient pipeline and memory access patterns. Thus, while W4A16 is promising for efficient inference, its practical speedup depends heavily on low-level kernel optimization and hardware-aware implementation.

\subsection{Ascend 910 NPU Architecture}
\label{subsec:ascend910}

The Huawei Ascend 910 NPU is a high-performance AI processor designed for both training and inference workloads, featuring a many-core architecture centered around specialized AI Cores. Each AI Core integrates dedicated compute, storage, and data movement units to accelerate matrix-intensive and vector-intensive operations prevalent in deep neural networks~\cite{ascend910}. Prior works on NVIDIA GPUs\cite{ampere} have explored similar W4A16 optimizations using Tensor Cores and shared memory tricks. However, the Ascend architecture's distinct memory hierarchy, instruction set, and lack of warp-level primitives necessitate a ground-up redesign—motivating our platform-specific kernel development\cite{frantar2025marlin}.

\begin{figure}[htbp]
  \centering
  \includegraphics[width=0.8\textwidth]{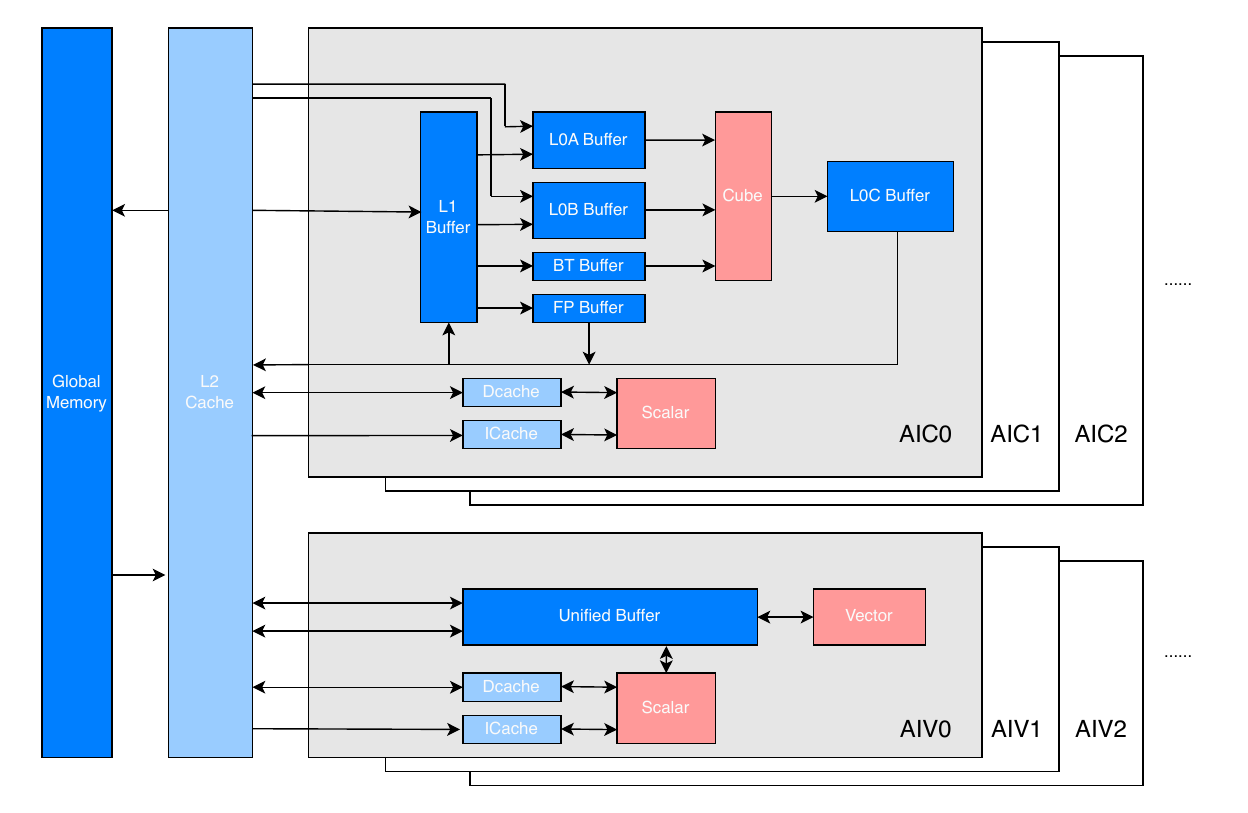}  % 替换为你的 PDF 路径
  \caption{Ascend NPU architecture.}
  \label{fig:npu}
\end{figure}

An AI core is organized around three primary components: compute units, the on-chip memory hierarchy, and Memory Transfer Engines (MTEs). As shown in Figure ~\ref{fig:npu}, the compute units consist of a high-throughput cube core (AIC) for matrix multiplication (e.g., FP16 GEMM with $16\times16\times16$ tiles), vector cores (AIV) for SIMD-like operations on data, and lightweight scalar cores handling control flow and instruction dispatch. Data reuse and minimized off-chip traffic are enabled by a dedicated on-chip memory hierarchy, which includes the L1 Buffer, the L0A/L0B input buffers, the L0C accumulation buffer, and the Unified Buffer (UB). Finally, high-bandwidth data movement between global memory (GM) and these on-chip buffers is managed by Memory Transfer Engines, which often perform on-the-fly data format conversions such as quantization or dequantization.The Ascend 910 operates in decoupled mode, the cube core and the vector core each operate under their own dedicated scalar scheduling units, which are separately deployed on the respective cores. An AI core is composed of one cube core and two vector cores on Ascend 910. 

Programming for Ascend 910 is typically performed via the Ascend C programming model\cite{CANN}, which abstracts the underlying hardware while exposing explicit control over data placement and pipeline scheduling. Efficient kernel implementation thus requires co-design of computation tiling, memory access patterns, and inter-unit synchronization to fully saturate the cube units—the primary throughput bottleneck in matrix multiplication workloads.

\section{Method}
We propose an efficient implementation of W4A16 mixed-precision matrix multiplication kernel in Huawei Ascend 910 NPU, leveraging the heterogeneous architecture comprising cube cores for high-throughput matrix compute and vector cores for flexible data processing. The algorithm adopts a Split-K\cite{cutlass-gemm} parallelization strategy to balance computation and memory bandwidth, which is particularly beneficial when the output matrix $C$ is small in dimension M or N (e.g., during the decode phase of LLMs).
Our kernel is inspired by the CATLASS\cite{CATLASS} Split-K implementation. Given input activation matrix $A \in \mathbb{R}^{M \times K}$ in FP16 and quantized weight matrix $W \in \texttt{INT4}^{K \times N}$ the goal is to compute:

\begin{equation}
\begin{aligned}
C = A \cdot (\texttt{Dequant}(W)) \\
\texttt{Dequant}(W) = s(W - z)
\end{aligned}
\label{eq:dequant_gemm}
\end{equation}

As shown in Algorithm~\ref{alg:Split-K_w4a16}, our method consists of three phases. First, in the dequant phase executed on vector cores, each vector core loads a slice of the quantized weight matrix $W$,  applies dequantization, unpacks INT4 to FP16, and writes the result to a global workspace buffer. The reason why dequantization is executed on vector cores is that cube cores do not support type conversion or element-wise multiply-accumulate operations. Second, the Split-K matrix multiplication phase is performed by cube cores, which compute partial matrix multiplications $C_i = A \cdot B_i$, where $B_i$ is the dequantized $i$-th $K$-slice, and accumulate the results into split buffers in global memory. Finally, in the Reduce phase, vector cores reduce the $S$ partial results via element-wise summation: $C = \sum_{i=0}^{S-1} C_i$, followed by type-cast from FP32 to FP16.

Marlin\cite{frantar2025marlin} finds that the GPU native type-cast is slow, and thus implements the INT4-to-FP16 data type-cast via PTX instructions with logical and multiply-accumulate (MAC) instructions. However, on the Ascend NPU, type-cast operations are exclusively executable on vector core. In this case, the bottleneck of W4A16 operator execution time no longer lies in the data type-cast itself, but in transferring the converted results back to cube core for subsequent computation. Therefore, we adopt the native data type-cast and hide the dequantization latency in data copy operations by leveraging double buffering strategy.

The entire pipeline exploits double buffering and hardware event-based synchronization between Memory Transfer Engines  and compute units to hide data movement latency. By leveraging the Split-K tiling strategy, the algorithm establishes a highly efficient pipeline on the Ascend NPU—spanning weight dequantization, tiled matrix computation, and result reduction—thereby significantly enhancing both computational throughput and energy efficiency for mixed-precision matrix multiplication.

\begin{algorithm}[htbp]
\caption{Split-K W4A16 matrix multiplication on Ascend NPU}
\label{alg:Split-K_w4a16}
\KwIn{
    $A \in \texttt{FP16}^{M \times K}$,
    $W \in \texttt{INT4}^{K \times N}$,
    split factor $S$,
    block size $[m, n, k]$,
    dequantization scale factor $s$,
    dequantization zero-point $z$
}
\KwOut{$C \in \texttt{FP16}^{M \times N}$}

\tcc{Phase 1: Dequant on vector cores (AIV)}
\ForPar{each vector core $v$}{
    \For{$i = v$ \KwTo $N/n$ \KwBy $\text{AIV}$}{
        \For{$j = 0$ \KwTo $K/k$ }{
        $W_{i,j} \gets W[j \cdot K/k:(j+1) \cdot K/k, i \cdot N/n:(i+1) \cdot N/n]$ \        \tcp*{Double buffer}
        $B_{i,j} \gets \texttt{Dequant}(W_{i,j})$ \tcp*{Dequantize to FP16}
        Write $B_i$ to workspace buffer at offset $i$ \tcp*{Double buffer}
        }
    }
}

\tcc{Phase 2: Split-K matrix multiplication on cube cores (AIC)}
\ForPar{each AI Core $c$}{
    \For{$i = 0$ \KwTo $S-1$}{
        \For{$l = 0$ \KwTo $M/m$ }{
            \For{$j = 0$ \KwTo $K/S/k$ }{
                \For{$r = c$ \KwTo $N/n$ \KwBy $\text{AIC}$}{        
                Load $A_{i,l,j}$ and $B_{i,j,r}$ from GM to on-chip buffers \ \tcp*{Double buffer}
                Compute $C_i^{(c)} = A_{i,l,j} \cdot B_{i,j,r}$ using \texttt{Mmad} \\
                Accumulate result to $C_i$ in global workspace
                }
            }
        }
        }
    }

\tcc{Phase 3: Reduction on vector cores (AIV)}
Wait for all AIC cores to finish matrix multiplication \\
\ForPar{each vector $v$}{
    Partition output elements among cores \\
    \texttt{Reduce()}
}
\end{algorithm}

\section{Experimental Results}
\label{sec:others}
\subsection{Kernel Strategy Benchmarking}
Unlike NVIDIA GPUs, which benefit from multiple open-source libraries such as Marlin, BitBLAS, and ExLlama\cite{exllamav2} for W4A16 mixed-precision matrix multiplication, the Ascend NPU currently lacks any native or optimized implementation of this operation. To address this gap, we implement a data-parallel W4A16 kernel based on CATLASS, which serves as the benchmark for comparison in this paper.

\begin{figure}[htbp]
    \centering
    % 第一行：3张图
    \begin{subfigure}[b]{0.33\textwidth}
        \centering
        \includegraphics[width=\linewidth]{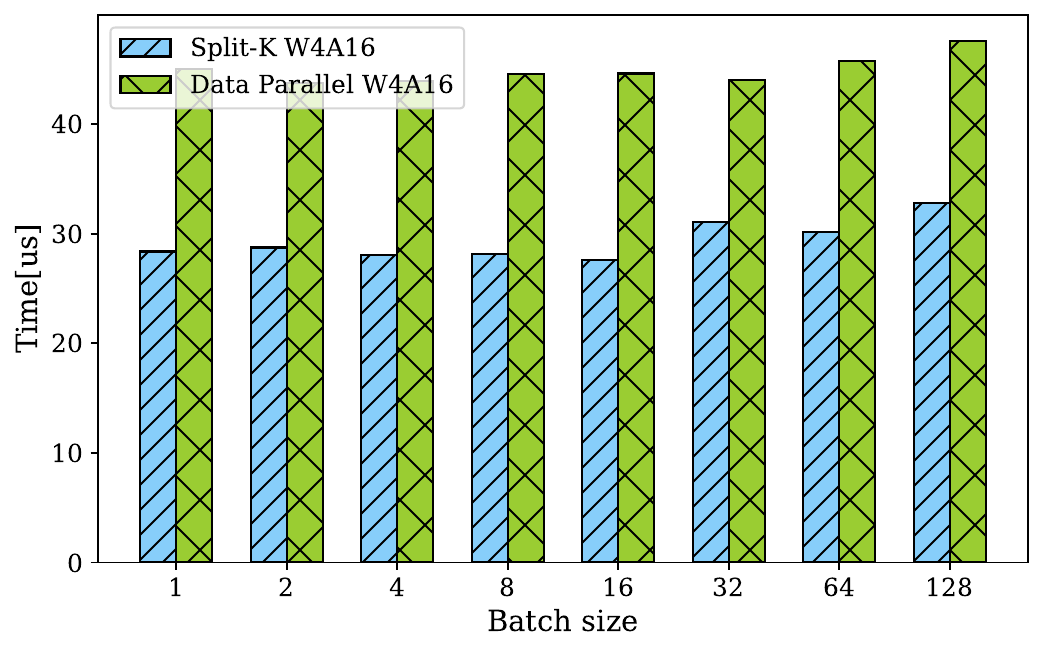}
        \caption{N=1536, K=6144}
        \label{fig:2a}
    \end{subfigure}
    \hfill % 水平填充，使子图分散
    \begin{subfigure}[b]{0.33\textwidth}
        \centering
        \includegraphics[width=\linewidth]{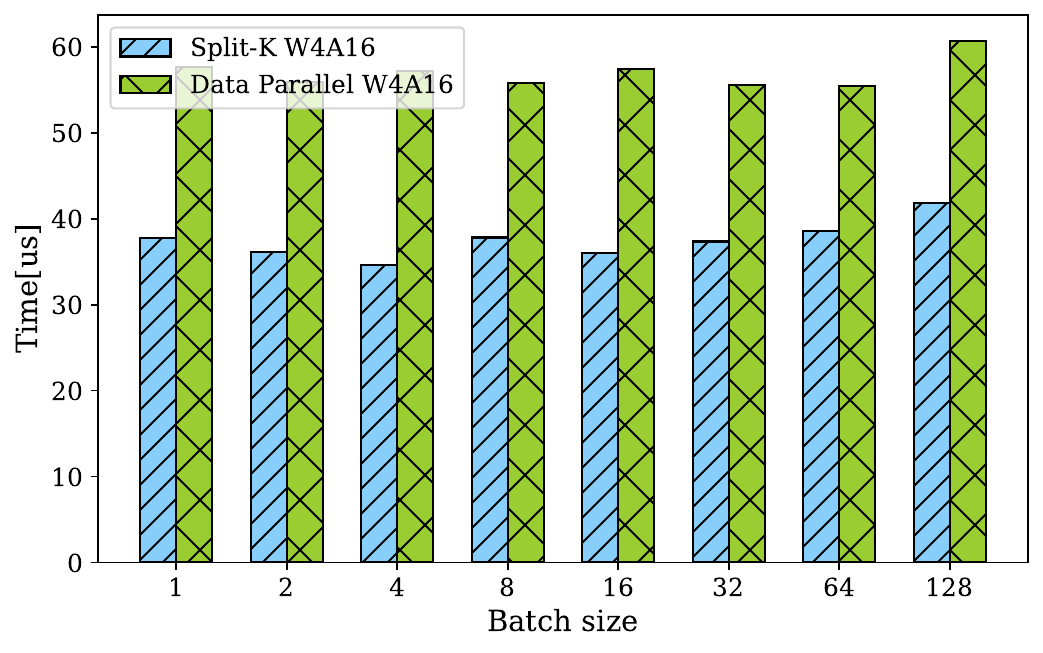}
        \caption{N=2048, K=8192}
        \label{fig:2b}
    \end{subfigure}
    \hfill
    \begin{subfigure}[b]{0.33\textwidth}
        \centering
        \includegraphics[width=\linewidth]{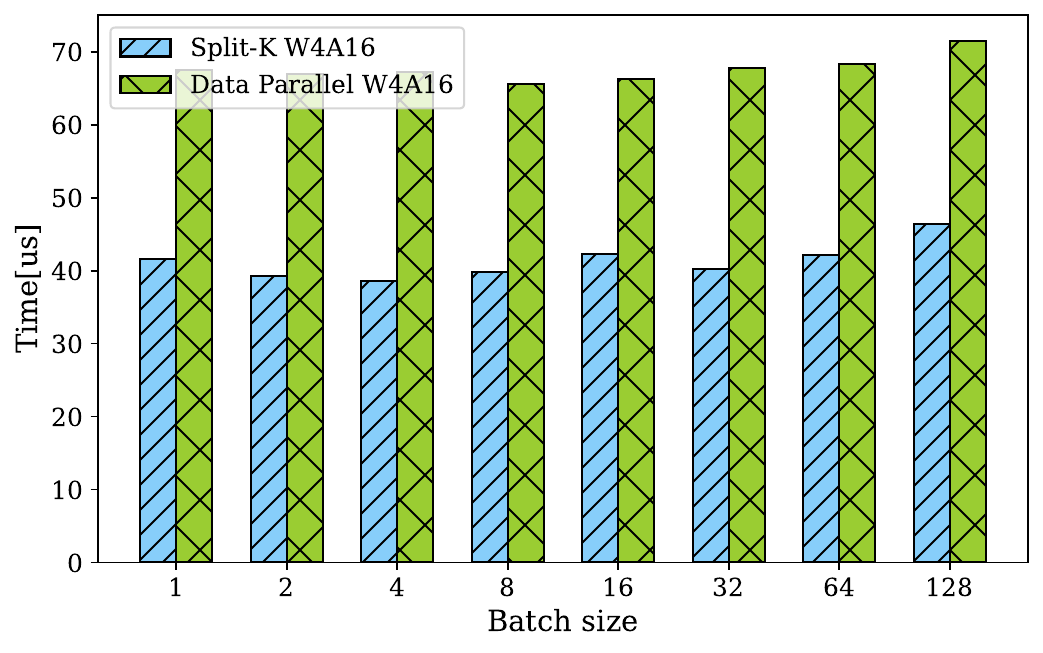}
        \caption{N=2048, K=10240}
        \label{fig:2c}
    \end{subfigure}
    
    \vspace{1em} % 行间距
    
    % 第二行：3张图
    \begin{subfigure}[b]{0.33\textwidth}
        \centering
        \includegraphics[width=\linewidth]{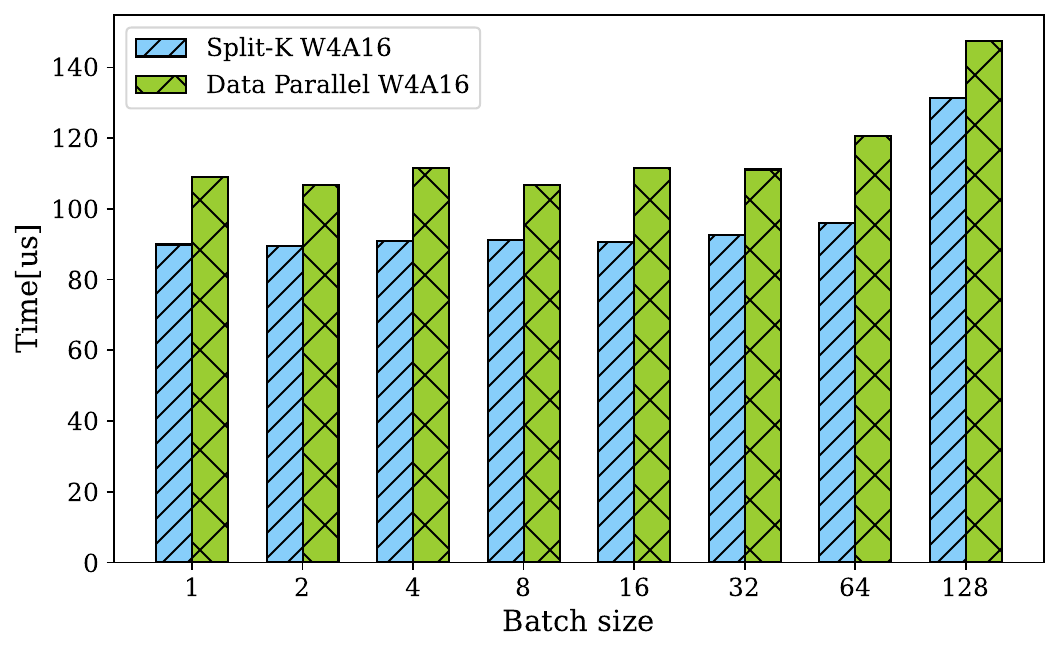}
        \caption{N=4096, K=16348}
        \label{fig:2d}
    \end{subfigure}
    \hfill
    \begin{subfigure}[b]{0.33\textwidth}
        \centering
        \includegraphics[width=\linewidth]{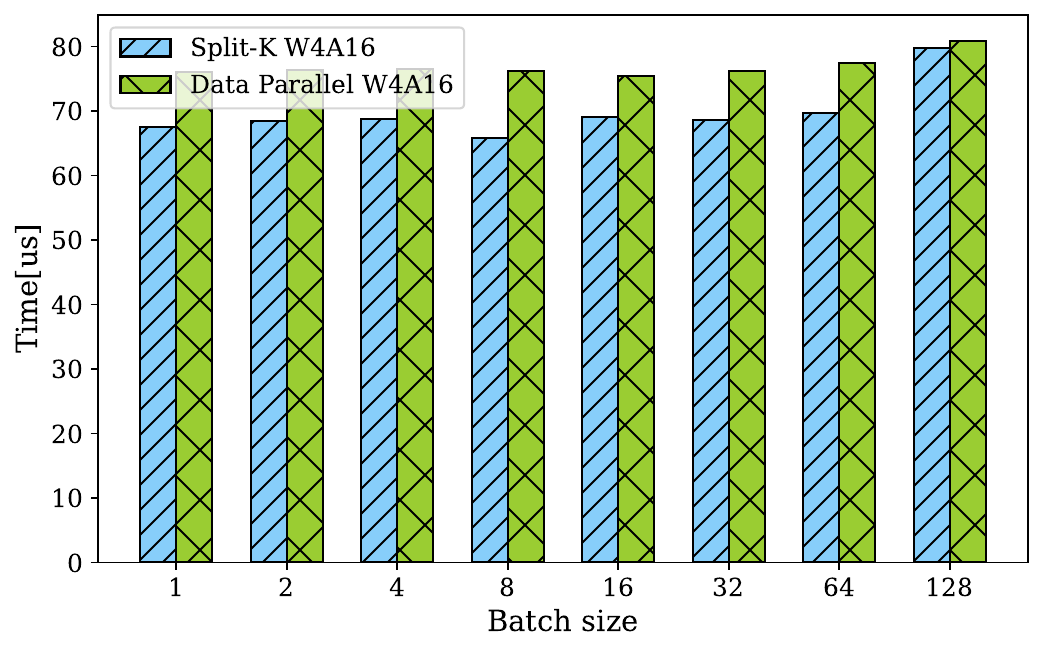}
        \caption{N=4608, K=10240}
        \label{fig:2e}
    \end{subfigure}
    \hfill
    \begin{subfigure}[b]{0.33\textwidth}
        \centering
        \includegraphics[width=\linewidth]{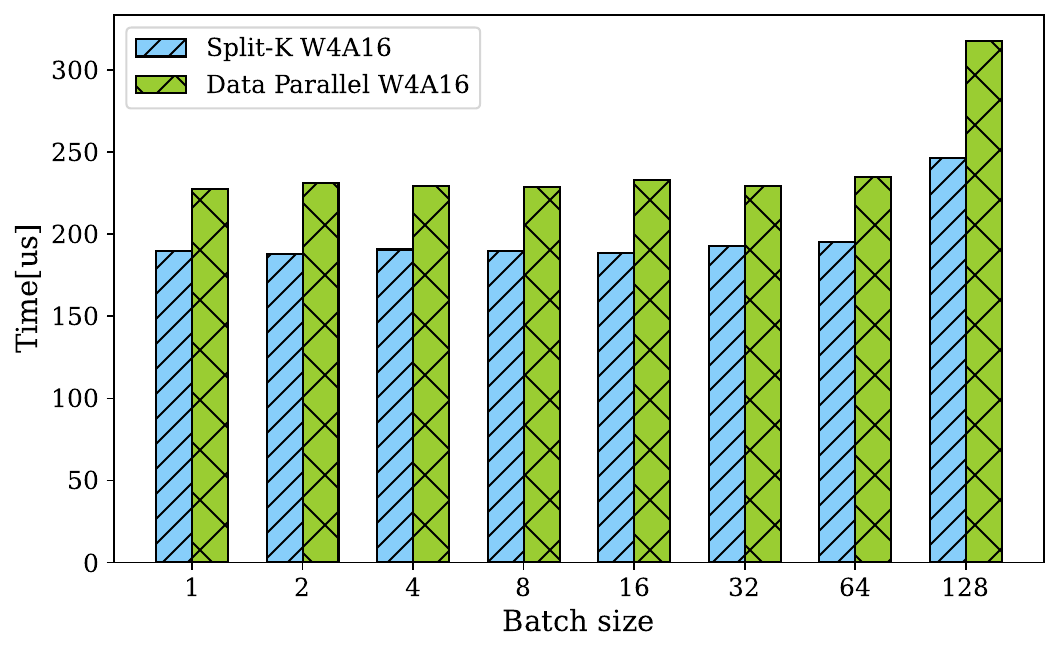}
        \caption{N=7168, K=18432}
        \label{fig:2f}
    \end{subfigure}
    
    \caption{Execution time of INT4-weight × FP16-activation matrix multiplication  on Ascend 910 NPU for various N×K configurations and batch sizes. The results compare two parallelization strategies: Split-K and Data Parallel.}
    \label{fig:2}
\end{figure}

We evaluate the performance of the INT4-weight × FP16-activation matrix multiplication kernel on the Ascend 910 NPU across a range of practical matrix dimensions derived from OpenPangu, DeepSeek-R1, GLM-4.5 and LLaMA3.2 as well as different batch sizes, with a comparative analysis of two parallelization strategies: Split-K and Data Parallel.
As shown in Figure~\ref{fig:2}, when $K$ is significantly larger than $N$, the Split-K strategy outperforms data-parallel approaches, and our method achieves a speedup ranging from $1.01\times$ to $1.74\times$. This is because Split-K can more effectively partition the computational workload across each cube core under such dimensional constraints. As the batch size (M) gradually increases, performance remains relatively stable. This occurs because the cube cores operate on matrix tiles as their fundamental processing granularity. Even when batch size is small, the input data is padded accordingly, which explains the limited acceleration benefit for small batch sizes. This behavior is observed in both Split-K and data-parallel methods.

\subsection{Memory Bottleneck of W4A16 Kernel}
\begin{figure}[htbp]
    \centering
    % 第一行：3张图
    \begin{subfigure}[b]{0.33\textwidth}
        \centering
        \includegraphics[width=\linewidth]{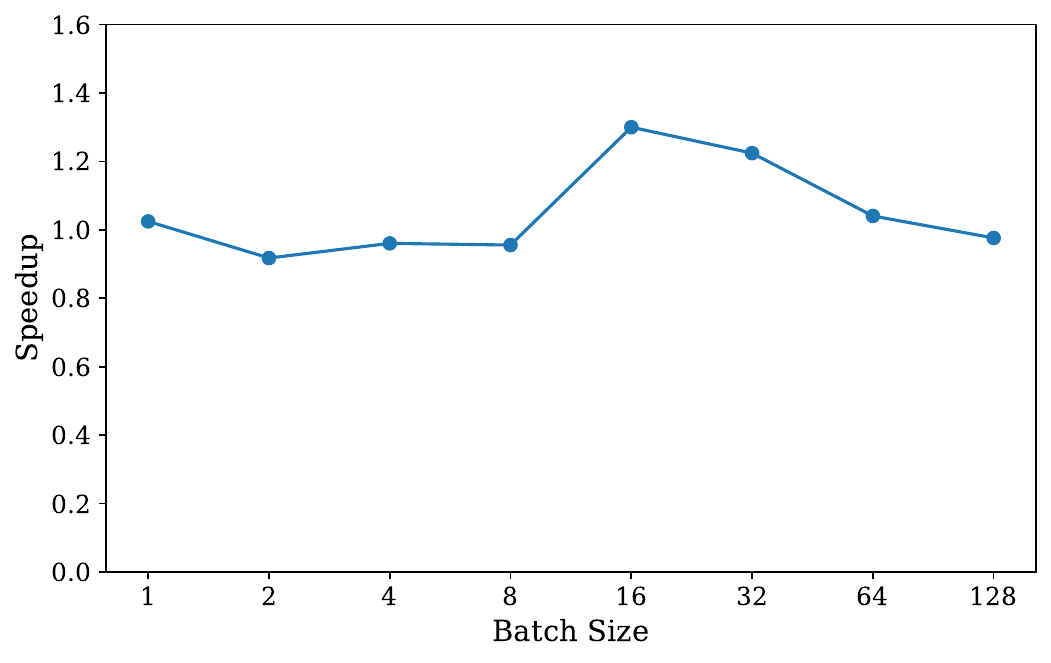}
        \caption{N=1536, K=6144}
        \label{fig:3a}
    \end{subfigure}
    \hfill % 水平填充，使子图分散
    \begin{subfigure}[b]{0.33\textwidth}
        \centering
        \includegraphics[width=\linewidth]{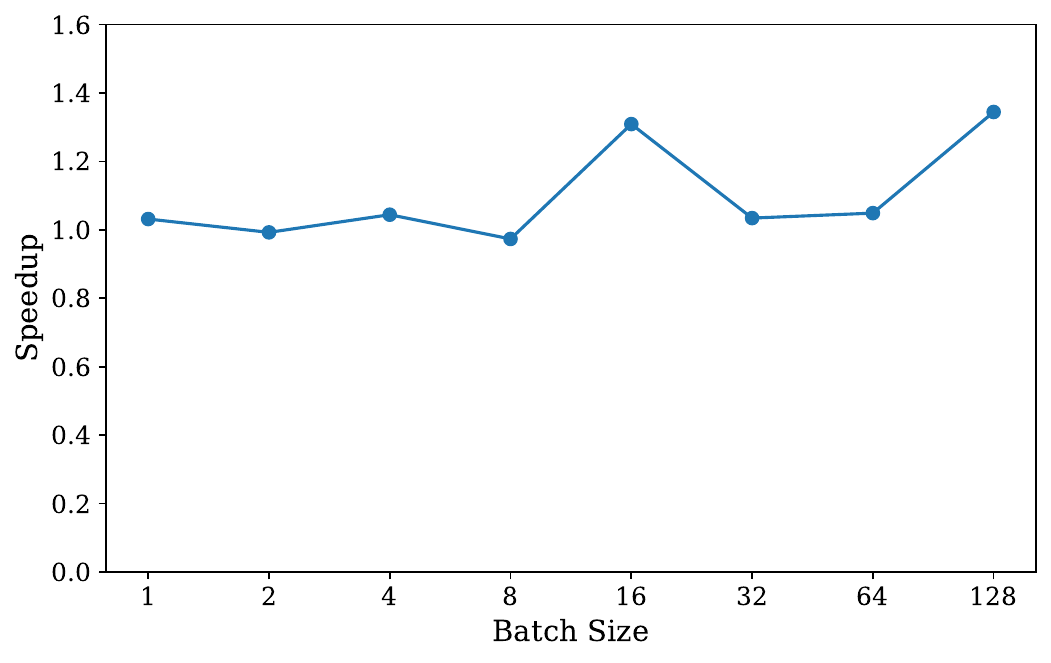}
        \caption{N=2048, K=8192}
        \label{fig:3b}
    \end{subfigure}
    \hfill
    \begin{subfigure}[b]{0.33\textwidth}
        \centering
        \includegraphics[width=\linewidth]{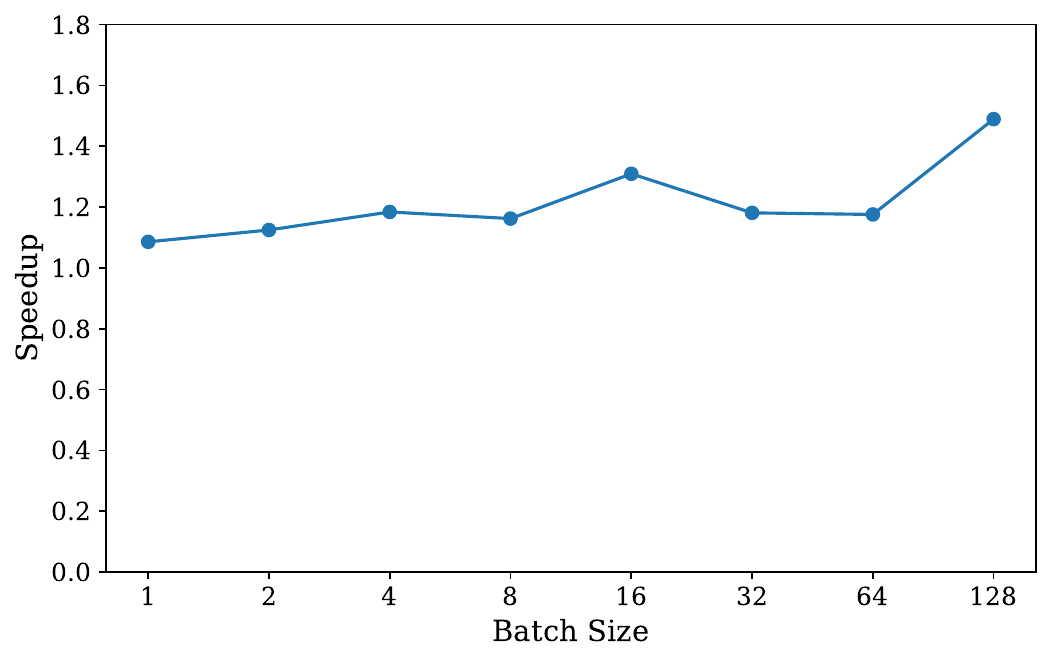}
        \caption{N=2048, K=10240}
        \label{fig:3c}
    \end{subfigure}

    \caption{Speedup of Split-K INT4×FP16 mixed-precision kernel over PyTorch' on Ascend 910 NPU for various N×K configurations and batch sizes.}
    \label{fig:3}
\end{figure}

The W4A16 kernel does not reduce the actual computational workload, it only alleviates global memory pressure by storing weights in 4-bit format. However, on the Ascend 910 NPU, the dequantization of weights must be performed by the vector core due to its decoupled architecture, while matrix multiplication is executed by the cube core. Since these two units communicate exclusively through global memory, the weight matrix must be written to global memory after dequantization and then read by the cube core. This incurs an additional memory round-trip compared to native FP16×FP16 matrix multiplication kernel. As a result, the W4A16 matrix multiplication does not achieve the theoretical $\sim$4$\times$ speedup compared to native FP16×FP16 matrix multiplication, especially at low batch sizes. Figure ~\ref{fig:3} corroborates our hypothesis that W4A16 matrix multiplication achieves at most a $1.48\times$ speedup over FP16×FP16 matrix multiplication on Ascend 910 NPU due to additional global memory traffic.

\section{Conclusion and Future Work}
In this work, we propose the first practical W4A16 mixed-precision matrix multiplication kernel tailored for Huawei's Ascend 910 NPU. Through the co-design of dequantization on vector cores and matrix multiplication on cube cores, coupled with the adoption of a Split-K parallelization strategy, our implementation delivers robust performance across a diverse range of matrix shapes – and achieves exceptional performance in the small-batch, large-K scenarios that are characteristic of LLM decoding. Compared with the data parallel scheme, our method yields a speedup of $1.01\times$ to $1.74\times$. Furthermore, our experimental analysis reveals a counterintuitive finding that challenges common assumptions: the primary performance bottleneck in W4A16 execution does not stem from the dequantization computation itself, but rather from the additional global memory round-trips incurred by transferring dequantized weights between the decoupled vector and cube processing units. This overhead prevents W4A16 from achieving the theoretical $\sim 4 \times$ speedup expected from its 4$\times$ reduction in weight storage footprint, only reaching a maximum speedup of $1.48\times$ over native FP16$\times$FP16 matrix multiplication in PyTorch.

These findings highlight a distinct trade-off for the current generation of Ascend NPUs: W4A16 is a viable strategy for maximizing memory capacity to fit larger models on a single device, but it does not yet yield the latency benefits observed on other platforms. Future work should explore hardware-software co-design to enable direct data paths between vector and cube units or fused instructions that bypass global memory, thereby unlocking the full latency potential of low-bit quantization on decoupled NPU architectures.

%Bibliography
\bibliographystyle{unsrt}  
\bibliography{references}

\end{document}